\documentclass[pre,a4paper,onecolumn,notitlepage,nofootinbib]{revtex4-1}
\usepackage{amsmath}
\usepackage{bbm}
\usepackage{bm}
\usepackage{latexsym}
\usepackage{graphicx}
\usepackage{subfigure}
\usepackage{amsfonts}
\newcommand{\uni}[1]{\,\mathrm{#1}}
\newcommand{\beq}{\begin{equation}}
\newcommand{\eeq}{\end{equation}}

\begin{document}

\title{\bf Exact Solution for Statics and Dynamics of Maximal Entropy Random Walk on Cayley Trees}

\author{J.K. Ochab}\email{jeremi.ochab@uj.edu.pl}

\author{Z. Burda}\email{zdzislaw.burda@uj.edu.pl}

\affiliation{Marian Smoluchowski Institute of Physics and 
Mark Kac Complex Systems Research Center \\
Jagiellonian University, Reymonta 4, 30-059 Krak\'ow, Poland}

\date{\today}

\begin{abstract}
We provide analytical solutions for two types of random walk: generic random walk (GRW) and maximal entropy random walk (MERW) on a Cayley tree with arbitrary branching number, root degree, and number of generations. For MERW, we obtain the stationary state given by the squared elements of the eigenvector associated with the largest eigenvalue $\lambda_0$ of the adjacency matrix. We discuss the dynamics, depending on the second largest eigenvalue $\lambda_1$, of the probability distribution approaching to the stationary state. We find different scaling of the relaxation time with the system size, which is generically shorter for MERW than for GRW. We also signal that depending on the initial conditions there are relaxations associated with lower eigenvalues which are induced by symmetries of the tree. In general, we find that there are three regimes of a tree structure resulting in different statics and dynamics of MERW; these correspond to strongly, critically, and weakly branched roots.

\medskip


\end{abstract}

\maketitle

\section{Introduction}

After the theory of Brownian motion and diffusive processes was formulated in the seminal works by Einstein \cite{Einstein} and Smoluchowski \cite{Smoluch}, random walk (RW) models, which stem from time or space discretization of these processes, have continuously attracted attention. The most celebrated ones include the Polya random walk on a lattice \cite{Polya} and its generalizations to arbitrary graphs. RW has been discussed in thousands of papers and textbooks in statistical physics, economics, biophysics, engineering, particle physics, etc., and still is an active area of research.

Mathematically speaking, RW is a Markov chain which describes the trajectory of a particle taking successive random steps.
For instance, in the case of Polya random walk, at each time step the particle jumps onto one of the neighboring nodes with equal probability.
Generalization of this process to any graph is what we call the ordinary or generic random walk (GRW).

Another kind of a RW, one that maximizes the entropy of paths and hence named maximal entropy random walk (MERW), has been investigated recently \cite{ZB1,ZB2}. The same principle of entropy maximization earlier led to the biological concept of evolutionary entropy \cite{Evolutionary1,Evolutionary2}.
It was also used in the problem of importance sampling where it served as an optimal sampling algorithm \cite{H}.
Now, MERW enters also the realm of complex networks \cite{MERW+CN1,MERW+CN2,MERW+CN3,MERW+CN4,MERW+CN5}. 
Its defining feature results in equiprobability of paths of given length and end-points, which means that if information is sent between two places, MERW makes it impossible to resolve which route the information has traveled. Another unprecedented feature of this RW is the localization phenomenon on diluted/defective lattices, where most of the stationary probability is localized in the largest nearly spherical region free of defects \cite{ZB1,ZB2}.
It has been illustrated with an interactive online demonstration \cite{BW}. In this paper, for the first time we show not only how stationary distributions of GRW and MERW differ but also how their dynamics differs on Cayley trees, for which the results are obtained analytically.

The paper is organized as follows: we begin with Sec. II defining GRW and MERW in general. In Sec. III we restrict our considerations to Cayley trees, for whose adjacency matrix we solve the eigenvalue problem by generalizing the
method given in \cite{Cayley}. The scheme presented there is utilized in Sec. IV, where we determine the eigenvector to the largest eigenvalue of the adjacency matrix, and then in Sec. V we generalize part of this result to eigenvectors associated with \mbox{next-to-leading} eigenvalues. Sec. VI presents the solution for eigenvalue problem of GRW transition matrix, repeating the order of arguments from Sec. III. Based on results from previous sections, Sec. VII describes stationary distributions of GRW and MERW on Cayley trees. Sections VIII and IX concern relaxation times of those two random walks, with general remarks in the former and particular results in the latter. Details concerning the solution of eigenproblems are to be found in Appendices A and B.

\section{Generalities}

Let us consider a discrete time random walk on a finite connected undirected graph. We are interested in a class of random walks with a stochastic matrix $\mathbf{P}$ that is constant in time. 
An element $P_{ij} \ge 0$ of this matrix encodes the probability that a particle being on a node $i$ 
at time $t$ hops to a node $j$ at time $t+1$. These matrix elements
fulfill the condition $\sum_j P_{ij} = 1$ for all $i$, which means
that the number of particles is conserved.
Additionally, let us assume that particles are allowed to hop
only to a neighboring node. This can be formulated as $P_{ij}\leq A_{ij}$, where $A_{ij}$ 
is the corresponding element of the adjacency matrix $\mathbf{A}$ of the graph: 
$A_{ij}=1$ if $i$ and $j$ are neighbors, and $A_{ij}=0$ otherwise.
The generic random walk (GRW) is realized by the following stochastic matrix:
\beq
P_{ij} = \frac{A_{ij}}{k_i} \ ,
\label{Porw}
\eeq
where $k_i = \sum_j A_{ij}$ denotes the node degree. The factor $1/k_i$ in the above formula
produces uniform probability of selecting one of $k_i$ neighbors of the node $i$. Clearly this choice maximizes entropy of neighbor selection
and corresponds to the standard Einstein-Smoluchowski-Polya random walk. 
The stationary state\footnote{A stationary state exists if a graph is not bipartite, but even for bipartite graphs a semi-stationary state can be defined
by averaging probability distribution over two consecutive time steps.} is given by $\pi_i = k_i/\sum_j k_j$. The other important type
of random walk, maximal entropy random walk (MERW), maximizes the entropy 
of random trajectories. In other words, one looks for a stochastic matrix that maximizes entropy for trajectories of given length and given end-points. This 
is a global principle similar to the least action principle. It leads to the following stochastic matrix:
\beq
P_{ij} = \frac{A_{ij}}{\lambda_0} \frac{\psi_{0j}}{\psi_{0i}},
\label{Pmerw}
\eeq
where $\lambda_0$ is the largest eigenvalue of the adjacency matrix 
$\mathbf{A}$ and $\psi_{0i}$ is the $i$-th element of the corresponding eigenvector $\vec{\psi}_0$. By virtue of the Frobenius-Perron theorem all elements of this vector are strictly positive, because the adjacency matrix $\mathbf{A}$ is irreducible. The stationary state of the stochastic matrix $\mathbf{P}$ is given by Shannon-Parry measure \cite{P}:
\beq
\pi_i = \psi^2_{0i} \ .
\eeq
The last formula intriguingly relates MERW to quantum mechanics.
Namely, $\psi_{0i}$ can be interpreted as the wave function of the ground state of 
the operator $-\mathbf{A}$ and $\psi^2_{0i}$ as the probability 
of finding a particle in this state \cite{ZB1,ZB2}. The two types, (\ref{Porw}) and (\ref{Pmerw}), of a random walk have in general completely different properties, although on a $k$-regular graph exceptionally they are identical.

The stochastic matrix is not symmetric in general, so it may have different
right and left eigenvectors:
\beq
\mathbf{P}\vec{\Psi}_\alpha = \Lambda_\alpha \vec{\Psi}_\alpha \quad , 
\quad
\vec{\Phi}_\alpha \mathbf{P} = \Lambda_\alpha \vec{\Phi}_\alpha \ .
\eeq
Throughout the paper, we consider left eigenvectors to be rows and right eigenvectors to be columns.
It can be easily seen that all the eigenvalues and eigenvectors
of the stochastic matrix $\mathbf{P}$
can be expressed in terms of eigenvalues $\lambda_\alpha$
and eigenvectors of $\vec{\psi}_\alpha$ of the adjacency matrix
$\mathbf{A}$:
\beq
\Lambda_\alpha = \frac{\lambda_\alpha}{\lambda_0} \ , \ 
\Psi_{\alpha i} = \frac{\psi_{\alpha i}}{\psi_{0 i}} \ , \
\Phi_{\alpha i} = \psi_{\alpha i} \psi_{0 i} \ .
\eeq
In particular, $\Lambda_0=1, \Psi_{0 i}=1, \uni{and}\ \Phi_{0 i}=\psi_{0 i}^2=\pi_{0 i} \uni{\ for \ all}\ i$. The spectral decomposition of $\mathbf{P}$ reads
\beq
P_{ij} = \sum_\alpha \Lambda_\alpha \Psi_{\alpha i} \Phi_{\alpha j} =
\sum_\alpha \frac{\lambda_\alpha\psi_{\alpha i}\psi_{\alpha j}}{\lambda_0} \frac{\psi_{0 j}}{\psi_{0 i}} \ .
\label{sd}
\eeq
Thus, clearly all properties of MERW are encoded in the spectral decomposition of the adjacency matrix of a given graph.
In what follows, we analyze the spectral properties of adjacency matrices for Cayley trees, derive the stationary state 
and dynamical characteristics of MERW on these trees, and compare them to GRW.

\section{Cayley tree}
\label{sec:Cayley}

\begin{figure}[bpt!]
	\centering
		\includegraphics[width=0.49\textwidth]{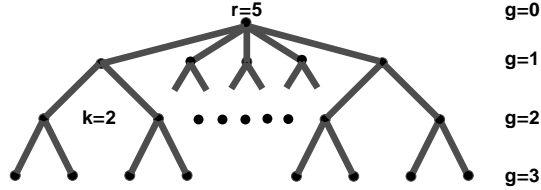}
	\caption{\label{fig:tree} Cayley tree with root degree $r=5$, branching number $k=2$, and $G=3$ generations.}
\end{figure}

Let us consider a Cayley tree with $G$ generations of nodes and
a branching number $k$ defined as the number of edges that connect a given node to nodes belonging to the next generation. We assume that the root of the tree has $r$ edges, which in general may be different from $k$ (see Fig. \ref{fig:tree}), and by convention, it belongs to the zeroth generation. Consequently, the zeroth generation contains one node, $n_0=1$, the first one $n_1=r$ nodes, the second one $n_2=rk$, the third one $n_3=rk^2$, and so forth. The total number of nodes in the tree is $n=\sum_{g=0}^G n_g =1+r(k^G-1)/(k-1)$.

The adjacency matrix of the underlying graph reads
\begin{equation}
\mathbf{A}=
\left( \begin{array}{ccccccc}
\mathbf{0} &\mathbf{B_0} & & & & &  \\
\mathbf{B^T_0} & \mathbf{0} & \mathbf{B_1} & & & &  \\
& \mathbf{B^T_1} & \mathbf{0} & \mathbf{B_2} & & &  \\
& &  & \ddots & \ddots & \ddots & \\
 & & & & & \mathbf{0} & \mathbf{B_{G-1}} \\
 & & & & & \mathbf{B^T_{G-1}}& \mathbf{0} \\
\end{array} \right),
\label{eq:adj}
\end{equation}
where the next-to-diagonal blocks $\mathbf{B_g}$ are rectangular matrices
of dimensions $n_g \times n_{g+1}$:
\begin{equation}
\mathbf{B_g}=
\left( \begin{array}{ccccccccccc}
1 &\ldots & 1 &   &             &        &            &   &        & \\
  &       &   & 1 & \ldots      & 1      &            &   &        & \\
  &       &   &   &             & \ddots & \ddots     &   &        & \\
  &       &   &   &             &        &            & 1 & \ldots & 1 
\end{array}\right) \ .
\end{equation}
Each line of $\mathbf{B_g}$ contains $k$ unities corresponding to branches leading to the descendent generation.
The block $\mathbf{B_0}$ reduces to a single-row matrix with $r$ unities. The matrices
$\mathbf{B_g^T}$ are the transposes of $\mathbf{B_g}$'s.

\subsection{Eigenvalues of the adjacency matrix}
\label{sec:adj}

In this section we calculate eigenvalues of the adjacency matrix
of Cayley tree using the method described in \cite{Cayley}.
The eigenvalues are given by solutions of the equation:
\beq
0=\det(\mathbf{A}-\lambda\mathbf{1}) =
\left| \begin{array}{ccccccc}
\mathbf{D_0} &\mathbf{B_0} & & & & &  \\
\mathbf{B^T_0} & \mathbf{D_1} & \mathbf{B_1} & & & &  \\
& \mathbf{B^T_1} & \mathbf{D_2} & \mathbf{B_2} & & &  \\
& &  & \ddots & \ddots & \ddots & \\
 & & & & & \mathbf{D_{G-1}} & \mathbf{B_{G-1}} \\
 & & & & & \mathbf{B^T_{G-1}}& \mathbf{D_G} \\
\end{array} \right|,
\label{eigen}
\end{equation}
where the diagonal blocks $\mathbf{D_g} = -\lambda \mathbf{1}$ are of
size $n_g \times n_g$, with $n_0=1$, $n_1=r$, and $n_g = rk^{g-1}$ for $g>1$. 
In order to calculate the determinant we use a sequence of elementary transformations such as additions of multiple of a row to another row, leaving the determinant invariant. This way the matrix is reduced to a triangular form with zeros above the diagonal.
First, we annihilate nonzero elements of the block $\mathbf{B_{G-1}}$ by multiplying rows that contain $-\lambda$ in the diagonal 
block $\mathbf{D_G}$ by $1/\lambda$ and adding them to the corresponding
rows in $\mathbf{B_{G-1}}$ that contain unities. This way all elements of $\mathbf{B_{G-1}}$ are turned to zero but at the same time the 
diagonal block $\mathbf{D_{G-1}}$ is modified to 
$\mathbf{D'_{G-1}}= -a_{G-1}\mathbf{1}$, where 
$a_{G-1} = -\lambda + k/\lambda$. 
Now, this procedure can be repeated to set the block $\mathbf{B_{G-2}}$ to
zero by multiplying rows that contain diagonal elements of  $\mathbf{D'_{G-1}}$ by $1/a_{G-1}$ and adding them to rows that contain 
unities in $\mathbf{B_{G-2}}$. While doing so, we see that the diagonal block $\mathbf{D_{G-2}}$ has been modified to $\mathbf{D'_{G-2}}= -a_{G-2}\mathbf{1}$, where $a_{G-1} = -\lambda - k/a_{G-2}$. Proceeding with this scheme recursively for 
the whole matrix we eventually obtain a triangular matrix determinant:
\beq
\det(\mathbf{A}-\lambda\mathbf{1}) =
\left| \begin{array}{ccccccc}
\mathbf{D'_0} & & & & & &  \\
\mathbf{B^T_0} & \mathbf{D'_1} & & &  &  \mathbf{0} &  \\
& \mathbf{B^T_1} & \mathbf{D'_2} &  & & &  \\
& &  & \ddots & \ddots & \ddots & \\
 & & & & & \mathbf{D'_{G-1}} & \\
 & & & & & \mathbf{B^T_{G-1}}& \mathbf{D'_G} \\
\end{array} \right|
\end{equation}
with diagonal blocks $\mathbf{D'_g} = a_g \mathbf{1}$ of size $n_g \times n_g$ 
whose coefficients are given by
\begin{align}
a_G&=-\lambda, \nonumber \\
a_g&=-\lambda-k/a_{g+1},\quad \uni{for} \ g=G\!-\!1,\ldots,1, \\
a_0&=-\lambda-r/a_{1} \ .\nonumber
\label{eq:rekMERW}
\end{align}
The diagonal coefficients  
$a_G(\lambda) = -\lambda$, $a_{G-1}(\lambda)=-\lambda-k/\lambda$,
$a_{G-2}=-\lambda - k/(-\lambda-k/\lambda)$, etc., 
are nested fractions in the argument $\lambda$. 
Hence, the equation (\ref{eigen}) for eigenvalues $\lambda$ 
takes the following form:
\beq
\prod_{g=0}^G \left[a_g(\lambda)\right]^{n_g} = 0 \ .
\label{proda}
\eeq
It is convenient to rewrite the left-hand side of the above
equation as a product of polynomials instead of fractions. 
There is a natural set of polynomials which can be constructed 
from $a_g$'s to this end:
\begin{eqnarray}
A_0(\lambda) & = & a_G  = -\lambda,  \nonumber \\
A_1(\lambda) & = & a_G a_{G-1}  =  \lambda^2 - k, \nonumber \\
A_2(\lambda) & = & a_G a_{G-1} a_{G-2} =  -\lambda( \lambda^2 - 2k), \nonumber \\
&\ldots& \label{polynoms} \\
A_g(\lambda) & = & -\lambda A_{g-1}(\lambda) - k A_{g-2}(\lambda), \quad \uni{for} \ g<G, \nonumber \\
A_G(\lambda) & = & -\lambda A_{G-1}(\lambda) - r A_{G-2}(\lambda). \nonumber
\end{eqnarray}
The recursive formula given above is derived
by noticing that $A_g = A_{g-1} a_{G-g} =$ $A_{g-1}(-\lambda-k/a_{G-g+1})=$ $-\lambda A_{g-1}~-~k A_{g-2}$. 
The exception is $g=G$, since then, in the last step the coefficient 
$k$ has to be replaced by $r$. 
Expressed in terms of polynomials $A_g$ the equation (\ref{proda}) reads
\beq
\prod_{g=0}^G \left[A_g(\lambda)\right]^{m_g} = 0,
\label{productA}
\eeq
where $m_G=1$ and $m_{G-g} = n_{g} - n_{g-1}$, for $g=1,2,\ldots,G$, or
equivalently $m_{G-1} = r-1$, $m_{G-g}=r(k-1)k^{g-2}$ for $g=2,3,\ldots,G$.
A simple analysis of the last equation shows that $A_g(\lambda)$ are 
polynomials of order $g+1$. Moreover, all odd order polynomials have
a root equal to zero. Later, we shall see that the equation 
$A_g(\lambda)=0$ has $g+1$
real roots and that if $\lambda$ is a root, $-\lambda$ also is.
The total number of real roots of equation (\ref{productA})
counted with degeneracy $m_g$ is $\sum_g (g+1) m_g = \sum_g n_g = n$, 
so Eq. (\ref{productA}) gives all $n$ eigenvalues of the adjacency matrix. The equation $A_0(\lambda)=0$ gives eigenvalues $\lambda=0$ with the degeneracy $m_G=r(k-1)k^{G-2}$, the equation $A_1(\lambda)=0$ gives eigenvalues $\pm\sqrt{k}$ with the degeneracy $m_G=r(k-1)k^{G-3}$, etc. It should be noticed that some eigenvalues may be solutions of $A_g(\lambda)=0$ for different $g$. For instance $\lambda=0$ is a root of $A_g(\lambda)=0$ for all even $g$, 
so the total degeneracy of the eigenvalue $\lambda=0$ is $\sum_g (2g+1) m_{2g}$. 

It turns out that the solutions of equations
$A_g(\lambda)=0$ can be found systematically. The polynomials $A_g(\lambda)$ for $g<G$ 
(\ref{polynoms}) can be written in a concise form using an 
auxiliary parameter $\theta$ (see Appendix \ref{appA}):
\beq
A_g = k^{(g+1)/2}\frac{\sin[(g+2)\theta]}{\sin\theta},
\label{Ag}
\eeq
where
\beq
\cos\theta = -\frac{\lambda}{2\sqrt{k}} \ .
\label{theta_lambda}
\eeq
It can be checked by inspection that these equations indeed reproduce
the polynomials (\ref{polynoms}). For example, for $g=0$ one retrieves
$A_0=\sqrt{k} \sin(2\theta)/\sin\theta=2\sqrt{k} \cos\theta = -\lambda$;
for $g=1$, $A_1= k\sin(3\theta)/\sin\theta=k[4(\cos\theta)^2-1] =\lambda^2-k$,
etc., in agreement with (\ref{polynoms}). The equation for $A_G$ can be obtained by combining the last equation
in (\ref{polynoms}) $A_G= -\lambda A_{G-1} - k A_{G-2}$ 
with the explicit form of $A_{G-1}$ and $A_{G-2}$ (\ref{Ag}), which yields
\beq
A_G=k^{(G-1)/2} \frac{k \sin [(G+2)\theta]+(k-r) \sin (G\theta)}{\sin\theta},
\label{AG}
\eeq
where $\theta$ is given by (\ref{theta_lambda}). 
When the root of the tree has $r=k$ neighbors (equal to the branching number of the tree), the last equation reduces to the one for remaining generations (\ref{Ag}).

The eigenvalues of the adjacency matrix can be determined by finding
values of the auxiliary parameter $\theta$ for which $A_g$ (\ref{Ag})
and $A_G$ (\ref{AG}) are zero and inserting these values to the
formula $\lambda = -2\sqrt{k} \cos\theta$ (\ref{theta_lambda}).
As can be seen, $A_g$ (\ref{Ag}) for $g<G$ is equal zero for $\theta \ne 0$
fulfilling the equation
\beq
\sin[(g+2)\theta]=0
\eeq
that has $g+1$ solutions
\beq
\lambda_{g,j}=2\sqrt{k} \cos\left(\frac{\pi j}{g+2}\right),
\quad \uni{for} \ j=1,\ldots, g+1.
\label{gj}
\eeq
Each eigenvalue in this series is $m_g$ times degenerated, as follows
from (\ref{productA}). The situation is slightly more complicated 
for $g=G$, since the equation $A_G=0$ amounts to an equation for $\theta$
\beq
k \sin [(G+2)\theta]+(k-r) \sin (G\theta) =0
\label{thetaG}
\eeq
that can be solved analytically only for $r=k$ or $r=2k$. In the first
case, exactly the same formula as for $g<G$ (\ref{gj}) is obtained
\beq
\lambda_{G,j}=2\sqrt{k} \cos\left(\frac{\pi j}{G+2}\right),
\quad \uni{for} \ j=1,\ldots, G+1,
\label{Gj1}
\eeq
while in the second one
\beq
\lambda_{G,j}=2\sqrt{k} \cos\left[\frac{\pi (j-1/2)}{G+1}\right],
\quad \uni{for} \  j=1,\ldots, G+1 \ .
\label{Gj2}
\eeq
For other values of $r$ one has to solve (\ref{thetaG}) numerically.
The largest eigenvalue of the adjacency matrix is
$\lambda_0=\lambda_{G,1}$. For $r=k$ it is equal 
\beq
\lambda_0 = \lambda_{G,1}=2\sqrt{k} \cos\left(\frac{\pi}{G+2}\right),
\eeq
while for $r=2k$
\beq
\lambda_0 = \lambda_{G,1}=2\sqrt{k} \cos\left(\frac{\pi}{2G+2}\right) \ .
\label{2Gplus2}
\eeq
For other values of $r$ the eigenvalue $\lambda_0$ can be
determined approximately as discussed in Appendix \ref{appB}. The solutions can be divided into three
classes with respect to values of $r$: the first class for 
$r \in (0,2k - 2k/G)$, the second one for $r \in (2k-2k/G,2k+2k/G)$, and the third one for $r \in (2k+2k/G,+\infty)$. In the large $G$ limit, i.e., 
for $G \gg 2k$ the second class reduces to a single integer value of $r=2k$ 
for which the solution is known (\ref{2Gplus2}). The first class corresponds
to the values $r < 2k$ for which the approximate solution reads
\beq
\lambda_0 = 2\sqrt{k} \cos \frac{\pi}{G+\delta},
\label{Gplusdelta}
\eeq
where  
\beq
\delta \approx \frac{2k}{2k-r} \ .
\eeq
as explained in Appendix \ref{appB}.
For the third class, $r>2k$, the equation (\ref{thetaG}) has no real 
solutions in the range $(0,\pi/(G+1))$ and the largest eigenvalue $\lambda_0$
is obtained from a purely imaginary solution for $\theta$. 
The corresponding equations change from trigonometric to hyperbolic.
For large $G$ the solution can be approximated by
\beq
\lambda_0 = \frac{2\sqrt{k}}{\sqrt{1-x^2}},
\label{lambdax}
\eeq
where
\beq
x = z \left[ 1 -2\left(\frac{1-z}{1+z}\right)^{G+1}\right]
\label{x}
\eeq
and
\beq
z= 1-\frac{2k}{r} \ .
\label{z}
\eeq
Again we refer the reader to Appendix \ref{appB} for details.
One sees that $x$ approaches $z$ exponentially as $G$ grows, so
for large $G$ one can substitute $x$ by $z$ in (\ref{lambdax}) to 
eventually obtain
\beq
\lambda_0 \approx \frac{r}{\sqrt{r-k}} \ .
\eeq
As can be seen the largest eigenvalue for trees with a strongly
branched root, $r>2k$, behaves differently as compared to trees with 
a weakly branched root, $r<2k$. This eigenvalue is now larger 
than $2\sqrt{k}$, while it was smaller in the previous case, 
it grows with $r$, and it is weakly dependent on $G$.

\section{The eigenvector to the leading eigenvalue}

In order to obtain the stationary state of MERW the largest eigenvalue $\lambda_0$ and the squared elements of the eigenvector $\vec{\psi}_0$ associated with this eigenvalue are needed:
\beq
(\mathbf{A}-\lambda_0\mathbf{1})\vec{\psi}_0=0.
\label{Al0}
\eeq
The ground state $\vec{\psi}_0$ has a helpful symmetry in the sense that all elements $\psi_{0i}$ for
nodes in a given generation $g$ are identical. 
So the problem can be simplified by ascribing the same value $\psi_g$ to all nodes
in the generation (henceforth, when we write out the elements of the eigenvector, we omit the index corresponding to the eigenvalue):
\beq
\vec{\psi}_0 = (\psi_{0},\underbrace{\psi_{1},\ldots,
\psi_{1}}_{n_1}, \
\ldots \ ,\underbrace{\psi_{G},\ldots, \psi_{G}}_{n_G}) \ .
\eeq
Effectively, instead of $n$ equations for $\psi_{0i}$, $i=1,\ldots,n$, 
(\ref{Al0}) there are just $(G+1)$ independent equations for $\psi_{g}$, $g=0,\ldots,G,$ left:
\begin{equation}
\begin{array}{rl}
-\lambda_0 \psi_{0} + r \psi_{1} & = 0, \\
\psi_{g-1} - \lambda_0 \psi_{g} +k \psi_{g+1} & =  0, \quad \uni{for} \ g=1,\ldots,G-2,  \\
\psi_{G-1} - \lambda_0 \psi_{G} & = 0.
\end{array}
\label{rekurencja}
\end{equation}
This recurrence can be solved starting from the end, $g=G$, and decreasing
$g$ to $0$. For convenience we introduce coefficients 
\beq
C_{g} =\frac{\psi_{G-g}}{\psi_{G}}
\label{Cgpsi}
\eeq
that invert the order of the recurrence. They correspond to
the original values normalized to $\psi_{G}$, in particular $C_0=1$.
The recurrence relations (\ref{rekurencja}) are equivalent to
\begin{equation}
C_g=\lambda_0 C_{g-1}-k C_{g-2} \ , \quad \uni{for} \  g = 2,\ldots,G,
\label{Cgr}
\end{equation}
with the initial condition $C_0=1$, $C_1=\lambda_0$. Let us note that 
the recurrence relation is identical as for $A_g$ (\ref{polynoms}) 
when $\lambda_0$ is replaced by $-\lambda_0$. The initial
condition is also identical, except that the counter of the recurrence 
is shifted by one, so the solution can be copied: $C_g(\lambda_0) = A_{g-1}(-\lambda_0)$ to obtain
\beq
C_g=k^{g/2} \frac{\sin[(g+1)\theta]}{\sin\theta}  \  , \quad \uni{for} \  g=0,\ldots,G,
\label{Cg}
\eeq
where $\cos\theta = \lambda_0/2\sqrt{k}$. The first equation in (\ref{rekurencja}) $-\lambda_0 \psi_0 + r \psi_{1} =0$, which corresponds
to an equation $-\lambda_0 C_G + r C_{G-1} = 0$, that is automatically fulfilled for $C_G$ and $C_{G-1}$ given by (\ref{Cg}) under substitution of $\lambda_0 = 2\sqrt{k}\cos\theta$ and $r \sin (G\theta) = k \sin [(G+2)\theta]+ k \sin (G\theta)$ according to the equation (\ref{thetaG}).

This concludes our calculations of the eigenvector to the leading
eigenvalue of the adjacency matrix. Using (\ref{Cgpsi}) we have
\beq
\psi_{g} = C_{G-g} \psi_{G} = \frac{C_{G-g}}{\sum_h C^2_h} 
\label{Psig}
\eeq
for all nodes in the $g$-th generation. The value $\psi_{G}$ is chosen to ensure the proper normalization $\sum_g \psi^2_{0,g}=1$.

\section{The eigenvector to next-to-leading eigenvalues}

In the case of the eigenvector $\vec{\psi}_1$ to the eigenvalue $\lambda_1$, we exploit the fact that it is symmetric within each of $r$ principal branches of the tree (which means that for given generation $g$ within the branch all the elements $\psi_{g}$ are the same; once again, when writing out the elements of the vector, we omit the index corresponding to the number of the eigenvalue). In appropriate coordinates, the elements belonging to these principal branches can be separated:
\beq
\vec{\psi}_1 = (\psi_{0},\alpha_1 \vec{\phi},\ldots,\alpha_r \vec{\phi}),
\eeq
where the branches may have different multiplicative factors $\alpha_1,\ldots,\alpha_r$ and the vector
\beq
\vec{\phi} = (\psi_{1}, \underbrace{\psi_{2},\ldots, \psi_{2}}_{n_2/r},\ldots, \underbrace{\psi_{G},\ldots,  \psi_{G}}_{n_G/r}),
\eeq
represents the relative value of the eigenvector elements in each branch. The multiplicities $n_g$ are evenly distributed among the $r$ branches, hence the factor $1/r$.

We obtain $(G+1)$ independent equations for $\psi_{g}$, $g=0,\ldots,G$, in analogy to the equation (\ref{rekurencja}):
\begin{subequations}

\begin{align}
-\lambda_1 \psi_{0} + (\alpha_1+\ldots+\alpha_r) \psi_{1} & = 0, \label{rekurencja1a}\\
\psi_{0}/\alpha_i - \lambda_1 \psi_{1} +k \psi_{2} & =  0, \quad \uni{for} \   i=1,\ldots,r,  \label{rekurencja1b}\\
\psi_{g-1} - \lambda_1 \psi_{g} +k \psi_{g+1} & =  0, \quad \uni{for} \  g=2,\ldots,G-1,  \label{rekurencja1c}\\
\psi_{G-1} - \lambda_1 \psi_{G} & = 0	.\nonumber
\end{align}

\end{subequations}
The only difference is the first two equalities above, which show how the $r$ branches couple together at the root of the tree. The rest of the equalities stay the same, as the recurrence progresses only within a given branch and the factor $\alpha_i$ is eliminated.

For each of the branches the system is solved starting from $g=G$ and decreasing $g$ to $1$. Until this point the solution is the same as before (\ref{Cg}).

Now, we check if (\ref{rekurencja1a}, \ref{rekurencja1b}) are consistent with this solution. Clearly, in equation (\ref{rekurencja1b}) the terms $-\lambda_0 \psi_{1} +k \psi_{2} = k^{G/2}\frac{\sin[(G+1)\theta]}{\sin\theta} \psi_{G} = 0$ , because $\lambda_1$ corresponds to the value $\theta = \frac{\pi}{G+1}$. Thus, after rewriting, the equations (\ref{rekurencja1a}, \ref{rekurencja1b}) take the form
\begin{subequations}
\begin{align}
\alpha_1+\ldots+\alpha_r & = 0 \label{alfy},	\\
\psi_{0}& =  0 .\label{}
\label{alpha}
\end{align}
\end{subequations}

In fact, $\psi_{0}= 0$ is consistent with the explicit solution $C_G\propto \sin[(G+1)\theta]=0$. If we recall the form of eigenvalues given in (\ref{gj}), of which one special case was $\lambda_1=\lambda_{G-1,1}$, it is noticeable that for each $g=0,\ldots,G-1$ the eigenvalue $\lambda_{g,1}$ corresponds to the angle $\frac{\pi}{g+2}$ and so the solution of the recurrence equation vanishes for generation $G-g$. This is the point at which the symmetry of the corresponding eigenvector is broken. Such a vector to the eigenvalue $\lambda_{g,1}$ has the elements $\psi_{g'}= 0$ for $g'<G-g$ and the symmetric values of $\psi_{g'}$ for $g'\geq G-g$. We do not discuss here the eigenvectors to the other eigenvalues.

The last point concerns the multiplication factors of branches $\alpha_1,\ldots,\alpha_r$. Noticeably, one of them can incorporate the normalization factor $\psi_G$, which leaves $r$ free parameters. There are however $r-1$ eigenvectors in the eigenspace of $\lambda_1$, so there are in fact $r(r-1)$ parameters [for lower eigenvalues one needs include the degeneration according to (\ref{productA})]. Now, there are also constraints: $r-1$ normalization conditions, $r-1$ constraints in (\ref{alpha}), and $\binom{r-1}{2}=\frac{(r-1)(r-2)}{2}$ pairwise orthogonalizations. This leads to the number
\beq
r(r-1)-(r-1)-(r-1)-\frac{(r-1)(r-2)}{2}=\frac{(r-1)(r-2)}{2}
\eeq
of free parameters, which are the allowed rotations $O(r-1)$ of the eigenspace.

To illustrate this with a simple example let us take $r=3$, which gives $r-1=2$ eigenvectors to $\lambda_1$:
\begin{align}
\vec{\psi} &= (0,\alpha_1 \psi_{1},\ldots,\alpha_3 \psi_{1}, \ \ldots \ ,\alpha_1 \psi_{G},\ldots,\alpha_3 \psi_{G}),\\
\vec{\phi} &=(0,\beta_1 \phi_{1},\ldots,\beta_3 \phi_{1}, \ \ldots \ ,\beta_1 \phi_{G},\ldots,\beta_3 \phi_{G}).
\end{align}
Two normalization conditions, for $\vec{\phi}$ and $\vec{\psi}$, rid the equations (\ref{alfy}) of two parameters:
\begin{subequations}
\begin{align}
\alpha_1+\alpha_2+1 & = 0,\\
\beta_1+\beta_2+1 & = 0,
\end{align}
\end{subequations}
while the orthogonalization $\vec{\phi}\cdot\vec{\psi}=0$ gives
\beq
\alpha_1\beta_1+\alpha_2\beta_2+1=0
\eeq
and finally the symmetric relation between the two vectors is obtained, leaving one free parameter that rotates them
\beq
2+\alpha_1+\beta_1+2\alpha_1\beta_1=0.
\eeq

\section{The eigenvalues of the GRW transition matrix}

Under the same procedure of transforming the determinant to the triangular form, as explained in Sec. \ref{sec:Cayley}, the transition matrix of generic random walk defined in (\ref{Porw}) leads to similar recurrence equations as in (\ref{eq:rekMERW})
\begin{align}
a_G&=-\lambda, \nonumber \\
a_{G-1}&=-\lambda- \frac{k}{k+1}\frac{1}{a_{G}}, \nonumber \\
a_l&=-\lambda-\frac{k}{(k+1)^2}\frac{1}{a_{l+1}}, \quad \uni{for} \ g=G\!-\!2,\ldots,1, \\
a_0&=-\lambda-\frac{1}{k+1}\frac{1}{a_{1}}. \nonumber
\label{eq:rekGRW}
\end{align}
In the last equality, the factor $r$ appears in the numerator and denominator, so it cancels out, and the equations remain $r$-independent.

We proceed as before and define
\beq
A_g(\lambda)=\prod_{j=0}^{g} a_{G-j}(\lambda), \quad \uni{for} \ g=0,\ldots, G
\eeq
and hence we get the recurrence relations
\begin{eqnarray}
A_g(\lambda) & = & -\lambda A_{g-1}(\lambda) - \frac{k}{(k+1)^2} A_{g-2}(\lambda), \quad \uni{for} \ g<G, \nonumber \\
A_G(\lambda) & = & -\lambda A_{G-1}(\lambda) - \frac{1}{(k+1)^2} A_{G-2}(\lambda).
\label{polynomsGRW}
\end{eqnarray}

The general solution for $g<G$ is (see Appendix \ref{appA} for details)
\beq
A_g(\lambda)=\left[\frac{k}{(k+1)^2}\right]^{(g+1)/2} \frac{\sin[(g+2)\theta]-k \sin(g\theta)}{\sin\theta}, \quad \uni{for} \ g=0,\ldots, G-1,
\label{AgGRW}
\eeq
where $\cos\theta=\frac{\lambda}{2}\sqrt{\frac{(k+1)^2}{k}}$.
$A_G(\lambda)$ can be found by inserting the above solution to Eq. (\ref{polynomsGRW}):
\beq
A_G(\lambda)= \left[\frac{k}{(k+1)^2}\right]^{(G+1)/2} \frac{(2k\cos\theta-1-k^2)\sin(G\theta)}{k \sin\theta}.
\label{AGGRW}
\eeq

Now, the eigenvalues of GRW transition matrix are determined by finding values of the auxiliary parameter $\theta$ for which $A_g$ (\ref{AgGRW}) and $A_G$ (\ref{AGGRW}) are zero. We first solve the equation for $g=G$, which factorizes into two parts:
\beq{}
2k\cos(2\theta)=1+k^2,
\eeq
whose solution is the largest eigenvalue of the transition matrix
\beq
\lambda_0=1,
\eeq
the second part being
\beq{}
\sin(G \theta)=0,
\eeq
which gives 
\beq
\lambda_{G,j}=2\sqrt{ \frac{k}{(k+1)^2}}\cos\left(\frac{\pi j}{G}\right), \quad \uni{for} \ j = 1,\ldots,G.
\eeq

For $g<G$ we obtain
\beq
	\sin [(g+2)\theta]=k \sin(g\theta),
	\label{thetaGGRW}
\eeq
which has the same form as equation (\ref{thetaG}), but with different coefficients. For $k>1$ in (\ref{thetaGGRW}), we enter the same range of parameters as for $r \in (2k+2k/G,+\infty)$ in Eq. (\ref{thetaG}), which means that the solution leading to the largest eigenvalue in a given series is imaginary. The corresponding equations change from trigonometric to hyperbolic. Under substitution $k=\frac{z+1}{1-z},\ z=\frac{k-1}{k+1}$, where $z$ was given in (\ref{z}), definition (\ref{x}) reads
\beq
x=\frac{k-1}{k+1}\left[1-2k^{-(g+1)}\right].
\label{xGRW}
\eeq

We are particularly interested in the second largest eigenvalue (corresponding to the series $g=G-1$). For large $G$ the solution is approximated by
\beq
\lambda_1= 2\sqrt{\frac{k}{(k+1)^2}} \frac{1}{\sqrt{1-x^2}}
\label{lambdaxGRW}
\eeq
and it can be easily seen that $\lambda_1$ exponentially approaches $\lambda_0=1$ for large $G$.

\section{Stationary states of GRW and MERW on Cayley trees}

As mentioned earlier, a stationary state for a random walk on a graph exists if the graph is not bipartite. In the case of bipartite graphs a semi-stationary state can be defined by averaging probability distributions over two consecutive steps (because even and odd times are independent) or by averaging the state over initial configurations.

The stationary state of GRW of is given by the linear dependence on the degree of the vertices
\beq
\pi_i = \frac{k_i}{\sum_j k_j}, \quad \uni{for} \ i=1,\ldots,n,
\eeq
so the distribution is flat (degree $k_i=k+1$) but for the root (degree $r$) and leaves (degree $1$). If we sum the probabilities over whole generations the exponential factor appears:
\beq
\Pi_g =n_g \pi_{i} = k^{g-1}\frac{k^2-1}{k^G-1}, \quad \uni{for} \ g=1,\ldots,G \ \uni{and} \ i \in g.
\eeq

\begin{figure}[hbtp!]
	\centering
		\includegraphics[width=0.49\textwidth]{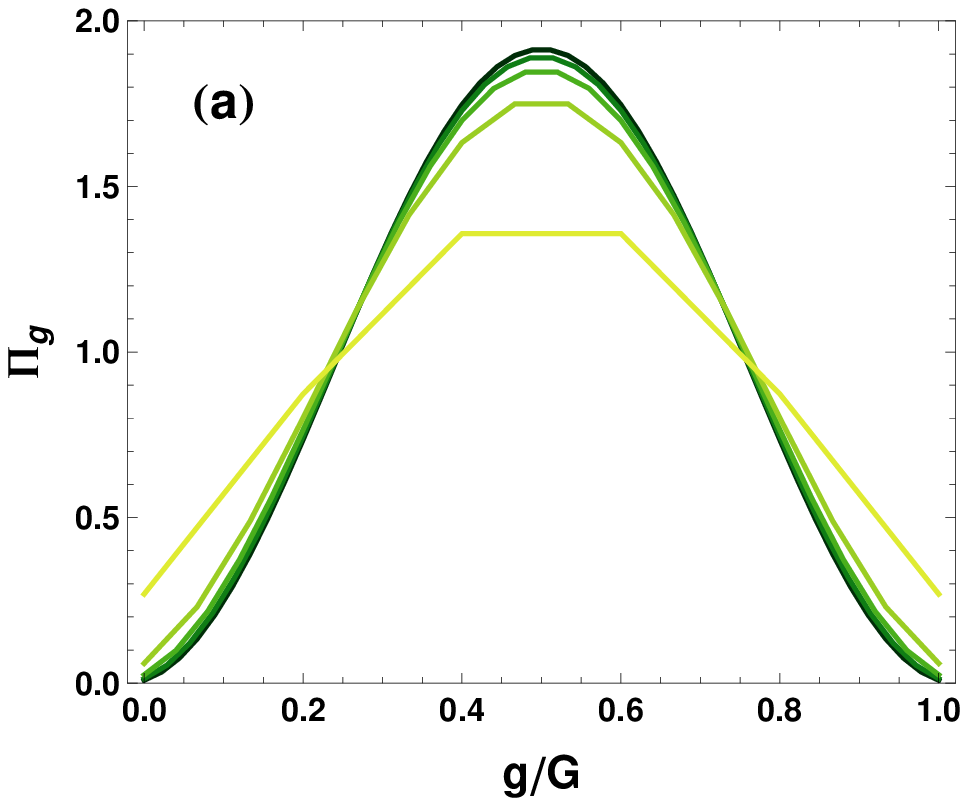}
		\hfill
		\includegraphics[width=0.49\textwidth]{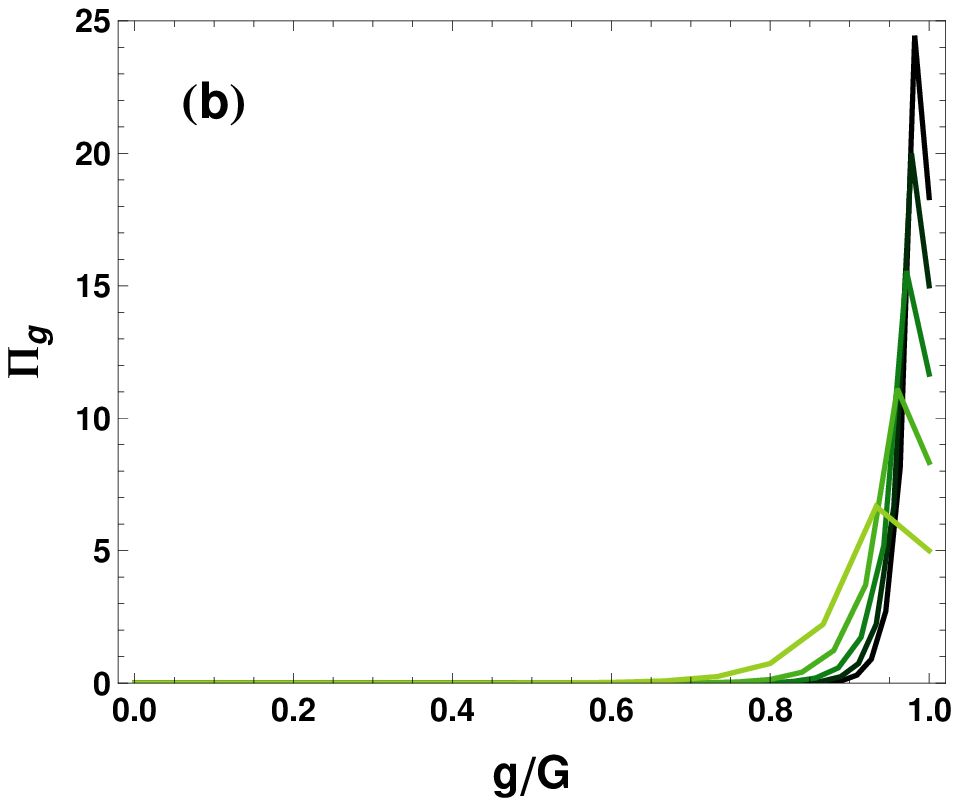}
	\caption{\label{fig:PGRW} Finite-size effect: the broken lines correspond to the distribution $\Pi_g$ for $G=5,\ldots,45$ in steps of $10$, for a tree with $k=r=3$. (a) The distributions for MERW. For $r<k$ the corresponding curves would be skewed and would approach the limiting distribution from  the right, while for $r>k$ from the left. (b) The distributions for GRW. 
	The larger the number of generations, the more peaked the distribution.}
\end{figure}

The stationary state of Maximal Entropy Random Walk is given by squared elements of $\vec{\psi}_0$, the eigenvector to the largest eigenvalue of the adjacency matrix:
\beq
\pi_i = \psi^2_{0i} \ .
\eeq
Remembering the solution (\ref{Psig})
\beq
\pi_{i}\propto k^{G-g} \sin[(G-g+1)\theta]^2, \quad \uni{for} \ g=0,\ldots,G \ \uni{and} \ i \in g,
\eeq
where we omitted the normalization factor. As we sum the stationary probability over $i\in g$ we get
\beq
\Pi_{g}=n_g \pi_{i}\propto k^{G-1} \sin[(G-g+1)\theta]^2, \quad \uni{for} \ g=1,\ldots,G \ \uni{and} \ i \in g,
\eeq
where the only exception is $g=0$ with its $n_0=1$. Exemplary probability distributions $\Pi_g$ for MERW and GRW are shown in Fig. \ref{fig:PGRW}.

\begin{figure}[hbtp!]
	\centering
		\includegraphics[width=0.49\textwidth]{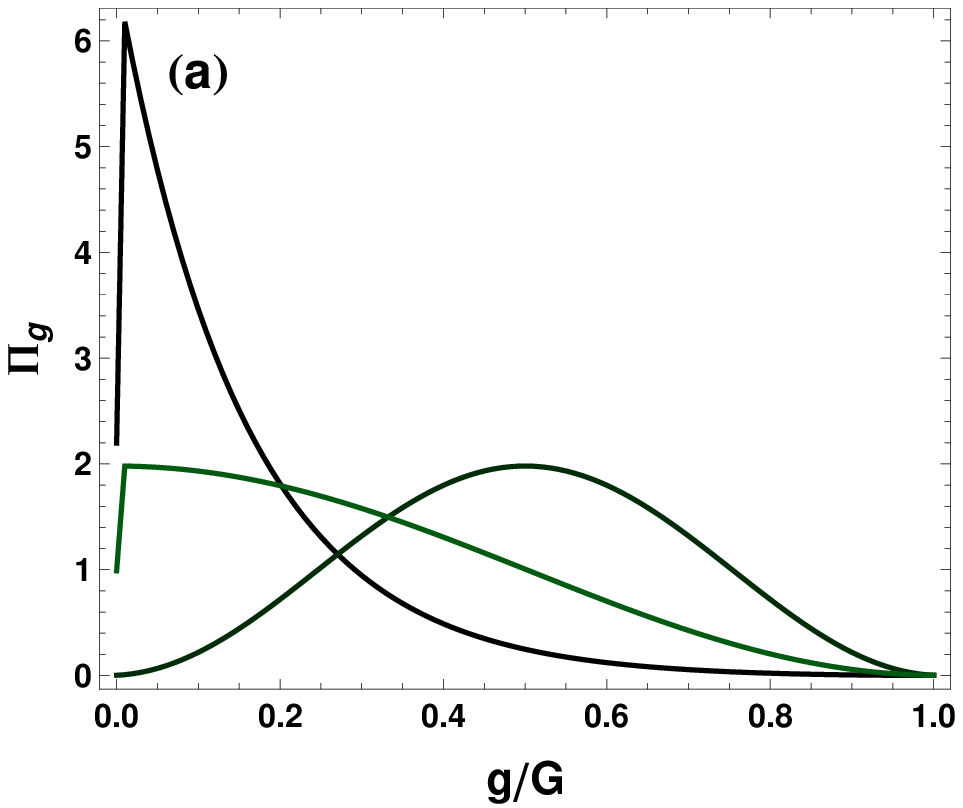}
		\hfill
		\includegraphics[width=0.49\textwidth]{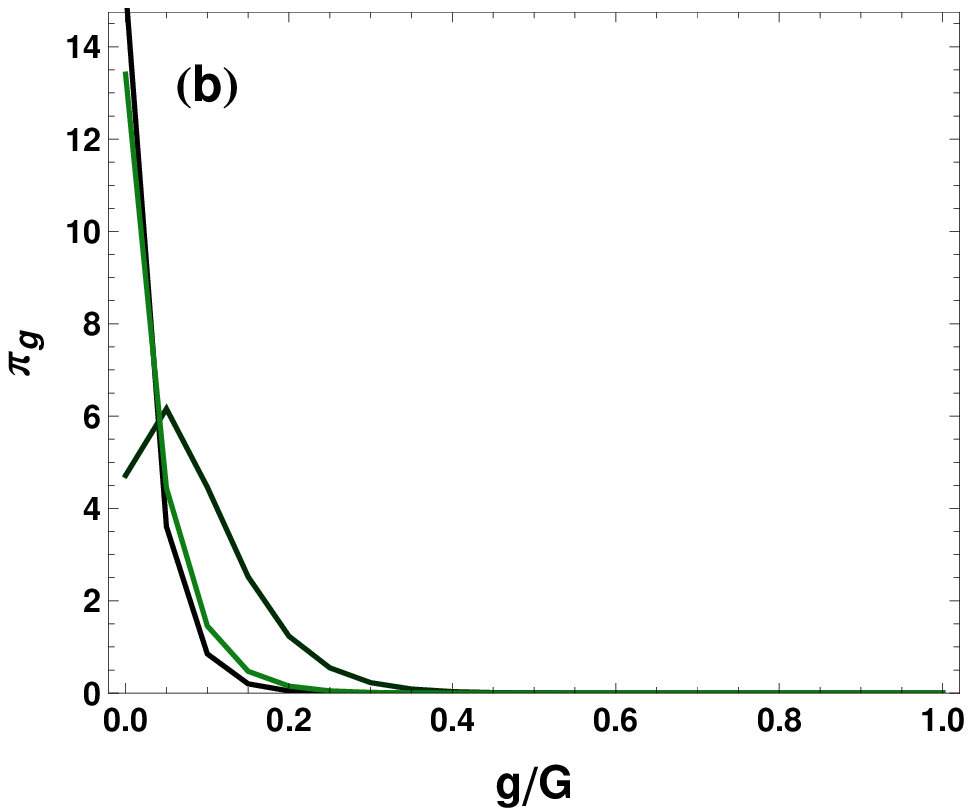}
	\caption{\label{fig:Psum}(a) Plots for $\Pi_g$, which is a total probability for a generation. Curves for large $G$: a sine square for $r<2k$, cosine square for $r=2k$, and hyperbolic sine for $r>2k$ ($k=3$ and $r=3,6,9$). (b) Probability per one node $\pi_g$ for $r=3,6,9$, $G=20$.}
\end{figure}

Now, as this result depends on $\theta$ and the solutions for $\lambda$ depend on whether $r<2k$, Eq. (\ref{Gj1}), $r=2k$, Eq. (\ref{Gj2}), or $r>2k$, Eq. (\ref{lambdax}), this means that we can get different distributions for different choices of $r$. For $r<2k$, parameter $\theta \approx \frac{\pi}{G+\delta}$ and the distribution remains a sine square; for $r=2k$, $\theta = \frac{\pi/2}{G+1}$ and the distribution becomes a cosine square; for $r>2k$, $\theta = i\ \uni{arctanh}\ x$ [where $x$ is given in (\ref{x}) and $i$ is the imaginary unit], thus we obtain a hyperbolic sine. Figure \ref{fig:Psum} illustrates these cases. An interactive demonstration showing these results as well as finite-size effects is available online \cite{Demo1}.

\section{Relaxation times}

\subsection{General considerations}
\label{sec:gen_relax}

Let us denote the probability of finding a particle at a node $i$ at time $t$ of
random walk by $\pi_i(t)$ and the probability distribution on the whole graph $\{\pi_i(t)\}_{i=1,\ldots,n}$ by $\vec{\pi}(t)$. Given the initial probability distribution $\vec{\pi}(0)$ and the stochastic matrix $\mathbf{P}$
one can determine the distribution at any time $t$
\beq
\vec{\pi}(t) = \vec{\pi}(0) \mathbf{P}^t
\eeq
Using the spectral decomposition of the stochastic matrix (\ref{sd})
one can rewrite the last equation as
\beq
\vec{\pi}(t) = \sum_\alpha c_\alpha \Lambda_\alpha^t \vec{\Phi}_\alpha \ .
\label{sd2}
\eeq
where $c_\alpha$ is a spectral coefficient: $c_\alpha=\vec{\pi}(0)\cdot \vec{\Psi}_\alpha = \sum_i \pi_i(0) \Psi_{\alpha i}$. In particular
$c_0 = \sum_i \pi_i(0) = 1$. In general all eigenvalues $\Lambda_\alpha$ of the stochastic matrix are known to be located inside or on the unit circle in the complex plane $|\Lambda_\alpha|\le 1$. In the limit $t\rightarrow \infty$ all terms in the sum on the right-hand side of the last equation for $|\Lambda_\alpha|<1$ are suppressed exponentially, and only
those for $|\Lambda_\alpha|=1$ survive. 
The stochastic matrices for GRW or MERW on trees have only two 
eigenvalues on the unit circle\footnote{More generally, since trees are bipartite one can show that if $\Lambda$ is an eigenvalue then also
$-\Lambda$ is.}: $\Lambda_0=1$ and $\Lambda_n=-1$, so for large $t$
one has 
\beq
\vec{\pi}(t) \approx c_0 \vec{\Phi}_0 + (-)^t c_n \vec{\Phi}_n .
\eeq
The eigenvectors associated with the eigenvalue $\Lambda_0=1$ are
$\Psi_{0i} = 1$ for all $i$, $\Phi_{0i} = \psi_{0i}^2=\pi_i$. In order to
write down the eigenvectors associated with the eigenvalue $\Lambda_n=-1$,
it is convenient to bipartition the graph into nodes belonging to generations numbered by odd and even $g$. Naturally, the ``odd" nodes are neighbors of ``even" ones only and vice versa. Elements of the eigenvectors are $\Psi_{n j_o} = 1$, $\Psi_{n j_e} = -1$,
$\Phi_{n j_o} = \pi_{j_o}$  and $\Phi_{n j_e} =-\pi_{j_e}$,
where the index $j_o$ runs over odd nodes and $j_e$ over even nodes.
This gives for large $t$
\beq
\begin{array}{llll}
\pi_{j_o}(2t) & = 2\sigma \pi_{j_o} &  ,  \ \pi_{j_e}(2t) &= 2(1-\sigma) \pi_{j_e} \\
\pi_{j_o}(2t+1) & = 2(1-\sigma) \pi_{j_o} &  , \  \pi_{j_e}(2t+1)&= 2\sigma \pi_{j_e}
\end{array}
\eeq
where $\sigma$ is the probability that a particle is in the odd part. Clearly, $c_n=2\sigma-1$ and for $\sigma=1/2$ the stationary state is recovered.
The equations above tell us that the probability distribution oscillates between odd and even nodes.
In a single step of a random walk particles disappear from odd nodes to appear on even ones and vice versa. If one traces the state of the random walk process every second step one sees that the distributions of particles on odd and even nodes approach the stationary state in each partition. The relaxation to the stationary state is generically governed by the \mbox{next-to-leading} eigenvalue
$\Lambda_1$ and its negative partner $\Lambda_{n-1}=-\Lambda_1$.
The corresponding term in the spectral decomposition (\ref{sd2}) reads
$\sum \left(c_1 \vec{\Phi}_1 + (-)^t c_{n-1} \vec{\Phi}_{n-1}\right) \Lambda_1^t$ and its contribution to the sum vanishes exponentially as $\exp(-t/\tau_1)$, where $\tau_1=[-\ln(\Lambda_1)]^{-1} = \left[\, \ln(\lambda_0/\lambda_1)\right]^{-1}$. The symbolic sum $\sum$ indicates that all eigenvectors in the eigenspaces of $\Lambda_1$ and $\Lambda_{n-1}$ are taken into account.
The exception is the case when the corresponding spectral coefficients $c_1$ and $c_{n-1}$ vanish, since then also the corresponding term vanishes.
In that case the \mbox{next-to-leading} contribution in the large $t$ limit comes from a lower eigenvalue $\Lambda_k$, the largest with a non-vanishing spectral coefficient. 

Thus, by $\tau_1$ we denote the generic relaxation time, the largest one, and by $\tau_2$ the one associated with $\lambda_{G,2}$ (in the Sec. \ref{sec:measure} we explain what symmetries lead to this relaxation). As there are several tree parameter regimes which yield different results for adjacency matrix eigenvalues, the relaxation times for MERW in those cases are also different. As explained in Appendix \ref{appB}, for eigenvalues of the adjacency matrix the relation $\lambda_{G-1,1}>\lambda_{G,2}$ always holds, so the second largest eigenvalue is $\lambda_1=\lambda_{G-1,1}$, unless some special parameters $k, r$ are chosen. Thus, we discuss below strongly, critically, and weakly branched root, and then some special cases. The discussion of relaxation for GRW and remarks on numerical measurements conclude this section. An interactive demonstration illustrating the results concerning relaxation is available online \cite{Demo2}.

\subsection{Strongly branched root}

The strongest root branching that yields qualitatively distinct behaviour of MERW is $r>2k$, where $k>1$ is assumed. The largest eigenvalue $\lambda_0$ is given by (\ref{lambdax}), while the second largest eigenvalue, with multiplicity $r-1$, belongs to the second level of hierarchy of eigenvalues
\beq
\lambda_1 = \lambda_{G-1,1}=2\sqrt{k}\cos\left(\frac{\pi}{G+1}\right).
\eeq
Thus, the generic relaxation time reads
\beq
\tau_{1}=-\left\{\ln\left[\sqrt{1-x^2} \cos\left(\frac{\pi}{G+1}\right)\right]\right\}^{-1},
\eeq
where $x$, defined in (\ref{x}), approaches $\frac{r-2k}{r}$ exponentially fast when $G\rightarrow \infty$. Hence, asymptotically:
\beq
\tau_{1}\cong c+\frac{c^2 \pi^2}{2}\frac{1}{G^2}+\ldots \longrightarrow c= \uni{const.},
\eeq
where
\beq
c=\left(\ln \frac{r}{2\sqrt{(r-k)k}}\right)^{-1},
\eeq
which gives an extremely fast relaxation, with the relaxation time converging to a constant for large $G$. A faster relaxation resulting from symmetry and associated with the eigenvalue $\lambda_{G,2}$ can be found as well, however the relaxation time might only be improved by a multiplicative constant.

\subsection{Critically branched root}

The behaviour of MERW changes for the special case of $r=2k,\ k>1$, as could be observed in the stationary states. The largest eigenvalue is given by (\ref{Gj2}) and the second largest eigenvalue $\lambda_1 = \lambda_{G-1,1}$ as before, hence the asymptotic relaxation time is
\beq
\tau_{1}\cong \frac{8 G^2}{3 \pi^2}+\frac{16 G}{3\pi^2}+\ldots.
\eeq
The symmetry-induced relaxation corresponding to $\lambda_{G,2}= 2\sqrt{k}\cos\left(\frac{3\pi/2}{G+1}\right)$, produces asymptotic behaviour with the same scaling with respect to the number of generations
\beq
\tau_{2}\cong \frac{1}{\pi^2}G^2+ \frac{2}{\pi^2}G+\ldots.
\eeq
It is worth noting that, while the number of vertices $n\sim k^G$, the probability distribution relaxes as a logarithm of the system size $\tau_{1},\tau_{2}\sim \ln n$, which still is rather fast.

\subsection{Weakly branched root}

After passing the critical value of $r=2k$ the tree enters the regime of weakly branched root, where $1<r<2k,\ k>1$. The only exact solution for $\lambda_0$ in this range of parameters is $r=k$ in (\ref{Gj1}), otherwise there is the approximation (\ref{Gplusdelta}) to our disposal. The second largest eigenvalue is the same as above, $\lambda_1=\lambda_{G-1,1}$. Hence, the generic relaxation follows
\beq
\tau_{1}\cong \frac{2k-r}{r \pi^2}G^3+\frac{3(4k-r)}{2 r \pi^2}G^2+\ldots
\eeq
and the faster relaxation relying on $\lambda_{G,2}$ gives
\beq
\tau_{2}\cong \frac{2}{3\pi^2}G^2- \frac{8k}{3\pi^2(r-2k)}G+\ldots \ .
\label{t2weakly}
\eeq
Noticeably, the generic relaxation time $\tau_{1}$ is $G$ times longer than $\tau_{2}$ and than both relaxation times for the tree with a critically branched root.

\subsection{Planted tree}

Until now we have considered only the root of degree $r>1$, where all the levels in the hierarchy of the eigenvalues have a non-zero degeneracy. Trees with a root of degree $r=1$ (known as \emph{planted trees}) are a special case, because the level $\lambda_{G-1,j}$ of the hierarchy has degeneracy $m_{G-1}=r-1=0$. Thus, the second largest eigenvalue is $\lambda_1 = \lambda_{G-2,1}$, while $\lambda_0$ is approximated by (\ref{Gplusdelta}) and the generic relaxation time is given by
\beq
\tau_{1}\cong \frac{2k-1}{2k\pi^2}G^3+ \frac{3}{2\pi^2}G^2+\ldots \ .
\eeq
The faster relaxation remains associated with the eigenvalue $\lambda_{G,2}$, so the asymptote (\ref{t2weakly}) is still valid for $\tau_{2}$ after inserting $r=1$.

\subsection{Linear chain}

Parameters $k=1,\ r=1$ produce a particularly degenerate case of a Cayley tree, namely a linear chain. While $m_{G-1} = r-1 = 0$ and $m_{G-g}=r(k-1)k^{g-2} = 0$, there remains only one level in the hierarchy of the eigenvalues of the adjacency matrix
\beq
\lambda_{G,j}= 2\sqrt{k}\cos\left(\frac{j \pi}{G+2}\right), \quad j = 1,\ldots,G+1.
\eeq
Naturally, $\lambda_{G,i} > \lambda_{G,j} \ \uni{for} \ \; i<j$, so $\lambda_0=\lambda_{G,1}$ and $\lambda_1=\lambda_{G,2}$, hence
\beq
\tau_{1}\cong \frac{2 G^2}{3\pi^2}+\frac{8 G}{3\pi^2}+\ldots
\eeq
and the relaxation connected with the third eigenvalue
\beq
\tau_{2}\cong \frac{G^2}{4\pi^2}+\frac{G}{\pi^2}+\ldots \ .
\eeq
However, if the number of generations $G$ is odd ($n$ even) there does not exist a central vertex, where this relaxation could be measured. If $G$ is even ($n=G+1$ is odd; it actually might be translated to $r'=2,k'=1,G'=G/2$ Cayley tree, although the solution differs from the previous ones) one central node exists and the faster relaxation can be measured there or if some symmetric initial conditions are taken.

Finally, let us notice that the system size is $n=G+1$ and the scaling is $\tau_1,\tau_2 \sim n^2$. This is the same result as for a simple diffusion, which is modeled by GRW.

\subsection{GRW relaxation times}

For GRW $\lambda_0=1$ and the second largest eigenvalue is given by (\ref{lambdaxGRW}) for all $k>1$. It follows that the relaxation time is given by
\beq
\tau_{1}\cong 2\left[\ln\left(1+ \frac{(k-1)^2}{4k} \frac{4k^{G}-1}{4 k^{G 2}} \right)\right]^{-1}.
\eeq
After using Taylor expansion, in the limit of large $G$
\beq
\tau_{1}\cong \frac{8k}{(k-1)^2} k^{G},
\eeq
which means that $\tau_{1}\sim n$.

The eigenvalue associated with the faster relaxation is $\lambda_{G,1}$ and it leads to the characteristic time
\beq
\tau_{2}\cong c-\frac{c^2\pi ^2}{2}\frac{1}{G^2}+ \ldots \longrightarrow  c= \uni{const.},
\eeq
where
\beq{}
c=-\left[\ln \left(2\sqrt{\frac{k}{(k+1)^2}}\right)\right]^{-1} .
\eeq

\subsection{Numerical measurements}
\label{sec:measure}

It is possible to measure the relaxation process in two ways: either explicitly taking powers of the transition matrix or Monte Carlo simulation with $N$ walkers traversing the graph.

In the former case: compute the transition matrix $\mathbf{P}$, choose the initial conditions (initial probabilities for any vertex of the graph), obtain the power of the transition matrix $\mathbf{P}^t$ (one might use spectral decomposition for that, although for large $t$ better precision is needed) corresponding to probabilities after $t$ steps, and measure the difference between the stationary state we have found theoretically. One might need to take the average of two consecutive steps to avoid the odd-even blinking.

In the case of Monte Carlo, let $N$ walkers start from node $a$ (or a set of nodes), every sweep for each of those walkers draw a random number and check it against the transition matrix to know in which direction the walker should go. At a node $b$ measure the number of random walkers at the sweep $t$, normalize it to the total number of walkers, and subtract the stationary state probability.

We have confirmed the theoretical relaxation times in both ways.

The difference between the the stationary state and the probability at time $t$ might be averaged over all nodes of the tree. However, to observe both the generic and the faster relaxation one might do one of the following:
\begin{enumerate}
	\item take one initial vertex with probability $1$, one measuring vertex,
	\item take $r$ initial vertices with probabilities $p_1, p_2,\ldots,p_r$, one measuring vertex.
\end{enumerate}
In the first case, if the initial vertex \emph{or} vertex at which one measures probabilities is the root, the observed relaxation time is $\tau_2$ and $\tau_1$ otherwise. In the second case, if the vertices and probabilities are chosen symmetrically (e.g., for $r=2$, the two neighbors of the root with probabilities $1/2$ each) one also sees $\tau_2$ if measuring the relaxation in the generation $g=1$. An interactive demonstration allowing to study this behaviour is available online \cite{Demo2}.

In general, one might spot other relaxations upon specific choices of initial conditions. This may be seen as eliminating contributions from given eigenvalues in the spectral decomposition of $\mathbf{P}$ (\ref{sd}), as explained in Sec. \ref{sec:gen_relax}. Intuitively, this is the same phenomenon as interference of waves, although we deal with probability waves here.

\section{Conclusions}

In this paper, we have analytically derived the form of the stationary state for GRW and MERW on Cayley trees, which shows that the stationary probability of the latter is centered around the root of a tree in contrast to the flat distribution of the former. The dynamics of the probability approaching to the stationary state have proven to be generically faster for MERW (logarithmic with respect to the system size) than for GRW (linear w.r. to the system size).

While Maximal Entropy Random Walk is defined so as to keep all paths of a given length between two given points equiprobable, it might be considered a process capable of hiding the route the information has travelled, e.g., on the Internet. The properties of stationary probability distribution of MERW have already been used to enhance centrality measures in complex networks \cite{MERW+CN2}. Considering the faster dynamics of MERW and the connection of eigenvalues of the adjacency matrix to the paths' statistics (which are a basis for a number of community detection algorithms \cite{F}), this type of random walk may prove useful in finding community structures on complex networks.

\begin{acknowledgments}
Project operated within the Foundation for Polish Science International Ph.D. Projects Programme co-financed by the European Regional Development Fund covering, under the agreement no. MPD/2009/6, the Jagiellonian University International Ph.D. Studies in Physics of Complex Systems.
\end{acknowledgments}

\appendix
\numberwithin{equation}{section}
\section{\text{Difference equations}}
\label{appA}

In this Appendix, we provide the reader with a detailed solution of the recurrence equations (\ref{polynoms}) resulting in

\begin{align}
A_g =&-\lambda A_{g-1}-k A_{g-2},\quad \uni{for} \  g<G,\\
A_G =&-\lambda A_{G-1}-r A_{G-2}.\nonumber
\end{align}

These difference equations can be solved with two initial conditions
\beq
\begin{array}{ccc}
A_0&=&-\lambda,\\
A_{-1}&=&1,
\end{array}
\eeq
where the first condition is found in (\ref{polynoms}) and the second condition is chosen so as to stay in agreement with the recurrence relation (indeed, $A_1 =-\lambda A_{0}-k A_{-1}=\lambda^2-k$).

The characteristic polynomial of this difference equation yields $\alpha^2 + \lambda \alpha+k=0$, resulting in $\alpha =\frac{1}{2}(- \lambda \pm i \sqrt{4 k- \lambda^2})$, and using the notation
\beq
\begin{array}{ccc}
\cos\theta&=&-\lambda/2\sqrt{k}, \\
\sin\theta&=&\sqrt{1- (\lambda/2\sqrt{k})^2},
\end{array}
\eeq
the general solution is obtained
\begin{equation}
\label{eq:rozwiazanie}
A_g=k^{(g+1)/2} [\alpha_1 \cos(g \theta)+\alpha_2 \sin(g \theta)], \quad \uni{for} \ g=0,\ldots, G-1.
\end{equation}
The first and second initial condition, respectively, lead to
\begin{equation}
\begin{array}{ccc}
\alpha_1&=&2 \cos \theta,\\
\alpha_2&=&\frac{\cos (2 \theta )}{ \sin\theta},
\end{array}
\end{equation}
after insertion of which the solution takes the form
\begin{equation}
A_g=k^{(g+1)/2} \frac{\sin(2\theta )\cos (\theta  g)+\cos (2\theta )\sin (\theta  g)}{\sin\theta }=k^{(g+1)/2} \frac{\sin[\theta(G+2)]}{\sin\theta },\ \uni{for}\ g<G.
\end{equation}
The last value, $A_G$ is calculated separately due to the root having degree $r$ that may be different from $k$:
\beq
A_G=k^{(G-1)/2} \frac{k \sin[\theta (G+2)]+(k-r) \sin (\theta  G)}{\sin\theta}.
\eeq

In the case of GRW, the recurrence equations are given by (\ref{polynomsGRW}). The solution proceeds analogously, however, due to different coefficients the initial conditions need to be adjusted accordingly:
\begin{equation}
\begin{array}{ccc}
A_0&=&-\lambda,\\
A_{-1}&=&k+1.
\end{array}
\end{equation}
The general form of the solution remains the same as given above but for the prefactor $k^{(g+1)/2}$ substituted with $\left[k/(k+1)^2\right]^{(g+1)/2}$. The first and second initial conditions give
\begin{equation}
\begin{array}{cc}
\alpha_1=&2 \cos \theta,\\
\alpha_2=& \frac{\cos (2 \theta )-k}{\sin\theta},
\end{array}
\end{equation}
which eventually lead to the solutions [(\ref{AgGRW}), (\ref{AGGRW})].

\section{\text{Trigonometric equations}}
\label{appB}

\begin{figure}[hbtp]
	\centering
		\includegraphics[width=0.49\textwidth]{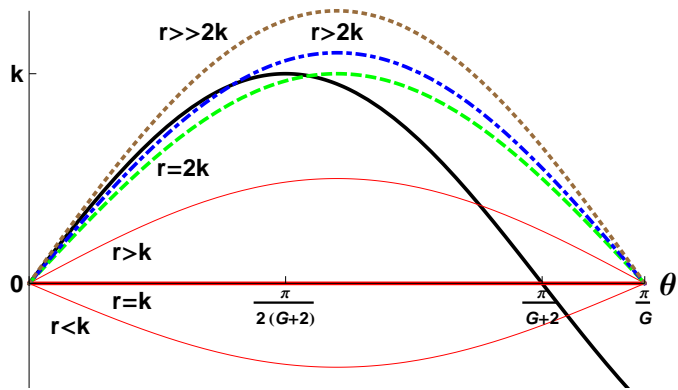}
		\hfill
	\caption{\label{fig:trig}(Color online) The intersection of the black curve with the other ones marks the solution of Eq. (\ref{thetaG}). The uppermost brown dotted curve corresponding to a strongly branched root shows no real solutions. The blue dot-dashed sine is an example of the rare case of strongly branched root with a real solution. The green dashed line is the critically branched root and the red continuous lines correspond to weakly branched roots.}
\end{figure}

In this Appendix, we derive more detailedly the approximate solutions to the trigonometric equations that appeared earlier in the article. The equation (\ref{thetaG}) can be illustrated with Fig. \ref{fig:trig}. For $r=k$ and $r=2k$ the analytical solutions (\ref{Gj1}) and (\ref{Gj2}) are found immediately. As mentioned in Sec. \ref{sec:adj}, for other values of $r$ the solutions can be divided into three classes with respect to values of $r$: the first class for $r \in (0,2k - 2k/G)$, the second one for $r \in (2k-2k/G,2k+2k/G)$, and the third one for $r \in (2k+2k/G,+\infty)$. In the large $G$ limit, that is for $G \gg 2k$ the second class reduces to a single integer value of $r=2k$ (although for small $G$ one can find several values, e.g., for $G=3,k=3,r=7$ the solution is still real). 

As regards the first class $r < 2k$, an approximation of the smallest $\theta$ (the largest $\lambda$) for large $G$ can be derived in the following way: let us transform equation (\ref{Gj1}) into
\beq
\tan\left[(G+1)\theta\right]=\frac{r}{r-2k}\tan \theta,
\eeq
In the limit $G\rightarrow\infty$ we expect $\theta \rightarrow 0$ (as we do observe such behaviour for $r=k$ and $r = 2k$), and upon Taylor expansion we obtain
\begin{subequations}
\begin{align}
\tan\left[(G+1)\theta\right] & \cong  \frac{r}{r-2k}\left(\theta+\theta^3/3\right),\\
(G+1)\left(\theta- \frac{\pi}{G+1} \right) & \cong \arctan\left[\frac{r}{r-2k}\left(\theta+\theta^3/3\right)\right],\\
(G+1)\left(\theta- \frac{\pi}{G+1} \right) & \cong  \frac{r}{r-2k}\theta+O(\theta^3).
\end{align}
\end{subequations}

Which, when having denoted by $\delta \approx \frac{2k}{2k-r}$, finally leads to
\beq
\theta  \cong \frac{\pi}{G+\delta}
\eeq
and produces the asymptotic solution (\ref{Gplusdelta}) for the first level of eigenvalues in the limit $G \rightarrow \infty$ for any branching parameters $k, r<2k$.

For the third class $r>2k$, the equation (\ref{thetaG}) has no real solutions in the range $(0,\frac{\pi}{G+1})$ and the largest eigenvalue $\lambda_0$
is obtained from a purely imaginary solution for $\theta$. The corresponding equations change from trigonometric to hyperbolic, so after transformation of (\ref{Gj1}) one gets
\beq
\tanh\left[(G+1)\theta\right]=\frac{r}{r-2k}\tanh \theta.
\eeq
For $G\rightarrow\infty$ this equation approaches
\beq
1=\frac{r}{r-2k}\tanh \theta^{*},
\eeq
which gives
\beq
\theta^{*}=\uni{arctanh}\left(\frac{r-2k}{r}\right).
\eeq
With the notation $z=1- \frac{2k}{r}$ and after utilizing the identity $\uni{arctanh}(z)=\frac{1}{2}\ln(\frac{1+z}{1-z})$
\beq
(G+1)\theta=\frac{1}{2}\ln\left(\frac{1}{z}\tanh\theta+1\right)-\frac{1}{2}\ln\left(1-\frac{1}{z}\tanh\theta\right).
\eeq
For large $G$ the first term on the right-hand side approaches $\frac{1}{2}\ln 2$, while the left-hand side $(G+1)\theta^{*}$. After rearranging this equation:
\beq
\theta\cong\uni{arctanh}\left\{z \left[1-\exp\left(\ln 2-2(G+1)\theta^{*}\right)\right]\right\},
\eeq
and finally under substitution of $\theta^{*}$:
\beq
\theta\cong\uni{arctanh}\left\{z\left[1-2\left(\frac{1+z}{1-z}\right)^{-(G+1)}\right]\right\}.
\eeq
The final solution (\ref{lambdax}) for $\lambda_0$ is due to the identity $\cos(i\ \uni{arctanh} x)=\frac{1}{\sqrt{1-x^2}}$.

The last remark concerns the problem of which eigenvalue $\lambda_{g,j}$ is the second largest one. If $r>1$, the level $G-1$ of the eigenvalue hierarchy exists. The eigenvalue $\lambda_{G-1,1}$ is defined by the angle $\theta_{G-1,1}=\frac{\pi}{G+1}$, whereas the second eigenvalue in the first level $\lambda_{G,2}$ is defined by an angle $\theta_{G,2}>\frac{\pi}{G}$. The latter information can be easily deduced from Fig. \ref{fig:trig}, where the intersections below the angle $\frac{\pi}{G}$ correspond to the largest eigenvalue. Thus, $\theta_{G-1,1}<\theta_{G,2}$ and consequently $\lambda_{G-1,1}>\lambda_{G,2}$. As this argument holds in general, $\lambda_1=\lambda_{G-1,1}$.

\bibliographystyle{plain}

\end{document}